\documentstyle[11pt,twoside,jltp,epsfig]{article}

\title{Interplay between Vorticity and Chirality inside the Vortex Core
in Chiral $p$-Wave Superconductors}

\author{Nobuhiko Hayashi\address{Computer Center, Okayama University,
Okayama 700-8530, Japan} and Yusuke Kato\address{Department of Basic Science,
University of Tokyo, Tokyo 153-8902, Japan}}

\runninghead{N. Hayashi and Y. Kato}{Interplay between Vorticity and Chirality
inside the Vortex Core \ldots}

\begin{document}

\begin{abstract}
   The vortex core in chiral p-wave superconductors
exhibits various properties
owing to the interplay between the vorticity and chirality
inside the vortex core.
   In the chiral p-wave superconductors,
the site-selective
nuclear spin-lattice relaxation rate $T_1^{-1}$
is theoretically studied
inside the vortex core
within the framework of the quasiclassical theory of superconductivity.
   $T_1^{-1}$ at the vortex center
depends on the sense of the chirality
relative to the sense of the magnetic field.
   The effect of a tilt of the magnetic field
upon $T_1^{-1}$
is investigated.
   The effect of the anisotropy in the superconducting gap and
the Fermi surface is then investigated.
   The result is expected to be experimentally observed as a sign of
the chiral pairing state in a superconducting material
Sr$_2$RuO$_4$.

PACS numbers: 74.60.Ec, 76.60.-k, 74.70.Pq
\end{abstract}

\maketitle

%Include this space if you do not use sections in your document.
%\vspace{0.3in}

% **********************************
\section{INTRODUCTION}
   Site-selective nuclear magnetic resonance (NMR) method
is a powerful tool for investigating
the electronic structure inside vortex cores
in the mixed state of type-II
superconductors.~\cite{Takigawa99,Matsuda02,Mitrovic01}
   We have theoretically studied~\cite{LT23haya,LT23kato}
the site-selective
nuclear spin-lattice relaxation rate $T_1^{-1}$
inside a vortex core in the case of
an isotropic
chiral $p$-wave superconductivity,~\cite{Sigrist99}
${\bf d}=
\bar{\bf z}(\bar{k}_{x} \pm i \bar{k}_{y})$.
   We found that
$T_1^{-1}$ was suppressed and almost vanished
in the $\bar{k}_{x} - i \bar{k}_{y}$ state
owing to the interplay between the vorticity and chirality
inside the vortex core
(here, the magnetic field was applied
in positive direction of the $z$ axis).~\cite{LT23haya,LT23kato}
   In this paper,
we
%extend our calculation to
%the case of inclined magnetic fields
%and
investigate the effect of a tilt of the magnetic fields
on $T_1^{-1}$.
   We then investigate
the effect of the anisotropy in the superconducting gap and
the Fermi surface.
   Our result is expected to be experimentally observed as a sign of
the chiral pairing state in a superconducting material
Sr$_2$RuO$_4$.~\cite{Sigrist99,Maeno}
%   However,
%our result~\cite{LT23haya,LT23kato} was
%in contrast to a corresponding theoretical result for $T_1^{-1}$
%calculated by Takigawa {\it et al.}~\cite{Takigawa02-2}
%   In their result,~\cite{Takigawa02-2}
%there was not
%such suppression of $T_1^{-1}$ as seen in ours.~\cite{LT23haya,LT23kato}
%   There can be two possible reasons for
%the difference between those results.
%   (i) The calculations of $T_1^{-1}$ by
%Takigawa {\it et al.}~\cite{Takigawa02-2}
%are in the quantum limit
%($k_{\rm F}\xi \sim 1$),
%while we base our calculations on
%the quasiclassical theory relevant in
%the opposite limit $k_{\rm F} \xi \gg 1$.~\cite{LT23haya,LT23kato}
%   (ii)

% **********************************
%\section{FORMULATION}
\section{QUASICLASSICAL THEORY}
   To investigate $T_1^{-1}$, we utilize
the quasiclassical theory of superconductivity.
%   The system is assumed to be a two dimensional conduction layer
%perpendicular to the magnetic field applied along
%the $z$ axis.
%   From now on, the notations are the same as those
%in Ref.\ ~\cite{Hayashi02}.
   We consider
the quasiclassical Green function
%
%================================
\begin{equation}
{\hat g}(i\omega_n,{\bf r}',{\bar{\bf k}})=
-i\pi
\pmatrix{
g &
if \cr
-if^{\dagger} &
-g \cr
},
\label{eq:qcg}
\end{equation}
%================================
which is the solution of the Eilenberger equation,~\cite{serene}
%================================
\begin{equation}
i {\bf v}_{\rm F}({\bar{\bf k}}) \cdot
{\bf \nabla}{\hat g}
+ \bigl[ i\omega_n {\hat \tau}_{z}-{\hat \Delta},
{\hat g} \bigr]
=0,
\label{eq:eilen}
\end{equation}
%================================
where
the superconducting order parameter is
${\hat \Delta}({\bf r}',{\bar{\bf k}}) =
\bigl[ ({\hat \tau}_{x} + i {\hat \tau}_{y}) \Delta({\bf r}',{\bar{\bf k}})
- ({\hat \tau}_{x} - i {\hat \tau}_{y}) \Delta^*({\bf r}',{\bar{\bf k}})
\bigr] /2$
and ${\hat \tau}_{i}$ the Pauli matrices.
${\bf v}_{\rm F}({\bar{\bf k}})$ is the Fermi velocity.
   The vector ${\bf r}'=(x',y',z')=(r\cos\phi,r\sin\phi,z')$
is the center of mass coordinate,
where the magnetic field is applied along
the $z'$ axis.
   The unit vector
${\bar{\bf k}}=(\bar{k}_{x},\bar{k}_{y},\bar{k}_{z})
%=(\cos\theta,\sin\theta, \bar{k}_{z})$
=({k}_{\perp} \cos\theta, {k}_{\perp} \sin\theta, {k}_{z})
/\sqrt{{k}_{\perp}^2 + {k}_{z}^2}$
represents
the wave number
of relative motion of the Cooper pairs
in the crystallographic coordinate frame.
%   We use units in which $\hbar = k_{\rm B} = 1$.

%   The overbar denotes unit vectors.
%   $\Delta({\bf r}',{\bar{\bf k}})$ is
%the superconducting order parameter.
%   $\omega_n$ is the Matsubara frequency.
%   The commutator $[{\hat a},{\hat b}]={\hat a}{\hat b}-{\hat b}{\hat a}$.
%   The cylindrical Fermi surface is assumed.
%and  $v_{\rm F}$ is the Fermi velocity.

   From the spin-spin correlation function,~\cite{Takigawa99}
we obtain the expression for $T_1^{-1}$
in terms of ${\hat g}$,
%================================
\begin{equation}
\frac{T_1^{-1}(T)}{T_1^{-1}(T_{\rm c})}
=
\frac{1}{4T_{\rm c}}
\int^{\infty}_{-\infty}
\frac{{\rm d} \omega}{\cosh^{2}(\omega / 2T)}
%\bigl(\frac{\omega}{2T}\bigr)}
%{\rm Re}
W(\omega, -\omega),
%                       \nonumber \\
%
%&  & \mbox{} +
%\frac{T}{\pi T_{\rm c}}
%%\int^{\infty}_{-\infty} {{\rm d} \omega}
%\int^{\infty}_{-\infty}
%{{\rm d} \omega}
%{{\rm d} \omega'}
%\mbox{\boldmath${\rm P}$}
%\frac{f_{\rm F}(\omega) - f_{\rm F}(\omega')}{(\omega - \omega')^{2}}
%   \nonumber \\
%&  & \mbox{} \times
%{\rm Im} W(\omega, -\omega'),
%
\label{eq:T1}
\end{equation}
%================================
%================================
\begin{equation}
W(\omega, \omega')
=
%\Bigl[
\langle a_{11}(\omega) \rangle
\langle a_{22}(\omega') \rangle
-
\langle a_{12}(\omega) \rangle
\langle a_{21}(\omega') \rangle,
%\Bigr],
\label{eq:T1-w}
\end{equation}
%================================
where the spectral function $\hat{a}(\omega,{\bf r}',{\bar{\bf k}})
=\bigl(a_{ij}\bigr)$ is given as
%================================
\begin{equation}
\hat{a}(\omega,{\bf r}',{\bar{\bf k}})
 = \frac{- i}{2 \pi} \hat{\tau}_3 \bigl[
%&  &
{\hat g}(i\omega_n \rightarrow \omega -i\eta ,
{\bf r}',{\bar{\bf k}})
%                             \nonumber \\
%&  & \mbox{} -
- {\hat g}(i\omega_n \rightarrow \omega +i\eta ,{\bf r}',{\bar{\bf k}})
\bigr],
\label{eq:spectral}
\end{equation}
%================================
the symbol $\langle \cdots \rangle$
represents
the average over the Fermi surface,
%\mbox{\boldmath${\rm P}$} means the principal part,
%$f_{\rm F} (\omega)$ is the Fermi distribution function,
and
$\eta$ is a small positive constant roughly representing
the impurity effect.
   $T$ is the temperature and
$T_{\rm c}$ the superconducting transition temperature.

% **********************************
%\section{RESULTS}
\section{EFFECT OF TILT OF MAGNETIC FIELD}
   We define $\gamma$ as the angle
between the magnetic field ($\parallel$ $z'$ axis)
and the $z$ crystallographic axis ($\parallel$ $\bar{k}_{z}$ axis).
   The isotropic cylindrical Fermi surface is assumed
and  the Fermi velocity is
${\bf v}_{\rm F}({\bar{\bf k}}) =
(v_{\rm F}\cos\theta,v_{\rm F}\sin\theta,0)$
in the crystallographic coordinate frame.
   We assume that the layer perpendicular to
the $z$ crystallographic axis is
isotropic.
   In this section,
we consider
the isotropic chiral $p$-wave pairing,~\cite{Sigrist99}
${\bf d}=
\bar{\bf z}(\bar{k}_{x} \pm i \bar{k}_{y})
= \bar{\bf z} \exp(\pm i\theta)$.
   We assume that this pairing
is unchanged by a tilt of the magnetic field.
%   The spin-spin correlation function,~\cite{Takigawa99}
%from which the expression for $T_1^{-1}$
%$\bigl[$Eqs.\ (\ref{eq:T1})--(\ref{eq:spectral})$\bigr]$
%is derived, is independent of the direction of the external magnetic fields.

   The order parameter around a vortex is expressed as
$\Delta({\bf r}',{\bar{\bf k}}) =
|\Delta(r)|\exp(i\phi) \exp(\pm i\theta)$.
   On the basis of an analysis of
the so-called zero-core vortex model,~\cite{Thuneberg84}
the matrix elements of
${\hat g}$ at the vortex center $r=0$
are
approximately obtained as~\cite{Hayashi02}
%%%%%%%%%%%%%%%%%%%%%%
%\begin{equation}
$g = \sqrt{\omega_n^2 +|{\tilde \Delta}|^2}
\omega_n^{-1}, \quad
f = -{\tilde \Delta} \omega_n^{-1}, \quad
f^{\dagger} = {\tilde \Delta}^{*} \omega_n^{-1}$,
%\label{eq:z-core-gf}
%\end{equation}
%%%%%%%%%%%%%%%%%%%%%
where
${\tilde \Delta}=
|\Delta (r \rightarrow \infty)|
\exp(i\phi) \exp(\pm i\theta)$
and
$(\cos \phi,\sin \phi) \parallel (v_{{\rm F}x'},v_{{\rm F}y'})$
in the plane perpendicular to the $z'$ axis, i.e.,
to the magnetic field.
   Without loss of generality,
we tilt the magnetic field in the $z$-$y$ plane.
   In the present case, it follows that
$\cos\phi=\cos\theta$ and $\sin\phi=\sin\theta \cos\gamma$.

%%%%%%%%%%%%%%%%%%%%%%%%%%%%%%%%%%%
\begin{figure}
%\vspace{-13mm}
%
\centerline{\psfig{file=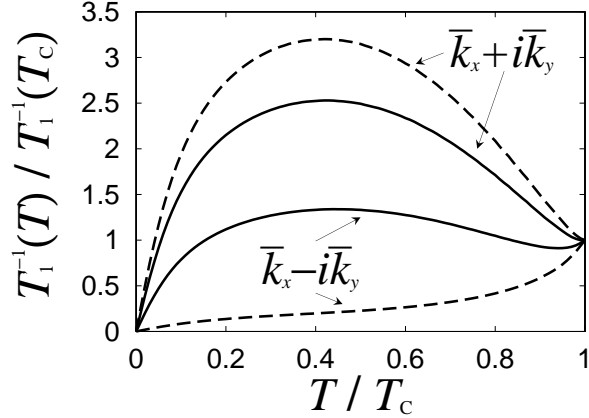,height=5.8cm}}
%
%\framebox[5in]{\rule[1.125in]{0in}{1.125in}}
%\makebox[5in]{\rule[1.125in]{0in}{1.125in}}
\caption{
   $T_1^{-1}$ vs $T$ at the vortex center.
   The magnetic field is tilted from
the positive direction of the $z$ crystallographic axis.
 The angle $\gamma$
between the magnetic field
and the $z$ crystallographic axis
is set to $\gamma=85^{\circ}$ (solid lines)
and $\gamma=0^{\circ}$ (dashed lines).
   The parameter $\eta = 0.2 \Delta_0$
($\Delta_0$ is the gap amplitude at $T=0$).
}
\label{fig:1}
\end{figure}
%%%%%%%%%%%%%%%%%%%%%%%%%%%%%%%%%%%

   Inserting these matrix elements of ${\hat g}$
into Eq.\ (\ref{eq:spectral})
and taking the Fermi-surface average
$\langle \cdots \rangle
=\int \cdots {\rm d}\theta /2\pi$ in Eq.\ (\ref{eq:T1-w}),
we numerically calculate $T_1^{-1}(T)$ at the vortex center.
   We show the result for $\gamma=85^{\circ}$ in Fig.\ \ref{fig:1}.
   Increasing the tilt angle $\gamma$,
we observe in our numerical results
that $T_1^{-1}(T)$ are almost unchanged by $\gamma$
up to $\gamma \sim 50^{\circ}$.
%   In $90^{\circ} > \gamma > 50^{\circ}$,
   Above $\gamma \sim 50^{\circ}$ ($\gamma < 90^{\circ}$),
each $T_1^{-1}(T)$
in the $\bar{k}_{x} \pm i \bar{k}_{y}$ states
gradually deviates from $T_1^{-1}(T)$ of $\gamma=0^{\circ}$
and the difference in $T_1^{-1}(T)$ between the two chiral states
gradually becomes small.
   At $\gamma = 90^{\circ}$, $T_1^{-1}(T)$
of the two chiral states
%in the $\bar{k}_{x} + i \bar{k}_{y}$ state
%and that in the $\bar{k}_{x} - i \bar{k}_{y}$ state
coincide each other.
   As seen in Fig.\ \ref{fig:1},
however,
even at a large tilt angle,
%$\gamma=85^{\circ}$,
the difference in $T_1^{-1}(T)$
between the $\bar{k}_{x} \pm i \bar{k}_{y}$ states
is still noticeable.
   At $\gamma=85^{\circ}$ (the solid lines in Fig.\ \ref{fig:1}),
   $T_1^{-1}(T)$ in the $\bar{k}_{x} + i \bar{k}_{y}$ state
is two times larger at $T \sim 0.4T_{\rm c}$
than that in the $\bar{k}_{x} - i \bar{k}_{y}$ state.

% **********************************
\section{EFFECT OF ANISOTROPY}

   Within the present framework
based on the quasiclassical theory,
we calculate $T_1^{-1}(T)$
using an anisotropic gap
${\bf d}=\bar{\bf z}(\sin k_x \pm i \sin k_y)$
and an anisotropic dispersion relation
$\varepsilon_k =
-2t(\cos k_x + \cos k_y)
-2t'\bigl[ \cos(k_x + k_y)+\cos(k_x - k_y)\bigr]$,
$\bigl($$t'=0.47t$ and
the chemical potential $\mu = 1.2t$~\cite{Takigawa02-1,Takigawa02-2}$\bigr)$.
   In this section, the magnetic field is applied
in positive direction of the $z$ crystallographic axis,
namely $\gamma=0^{\circ}$.
   This physical situation is the same as
that of a calculation of $T_1^{-1}(T)$
by Takigawa {\it et al.}~\cite{Takigawa02-2}

%%%%%%%%%%%%%%%%%%%%%%%%%%%%%%%%%%
\begin{figure}
%\vspace{-13mm}
%
\centerline{\psfig{file=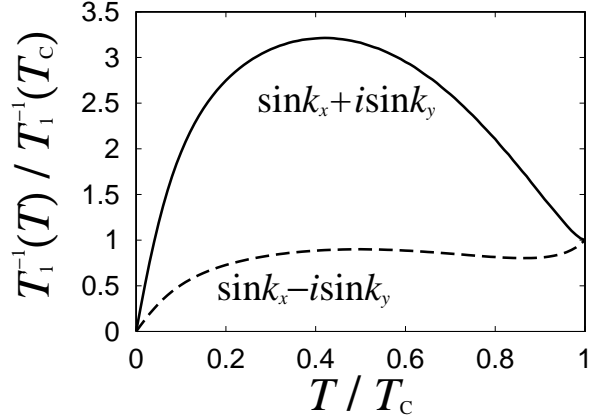,height=5.8cm}}
%
%\framebox[5in]{\rule[1.125in]{0in}{1.125in}}
%\makebox[5in]{\rule[1.125in]{0in}{1.125in}}
\caption{
   $T_1^{-1}$ vs $T$ at the vortex center.
   The magnetic field is applied
in positive direction of the $z$ crystallographic axis.
   The parameter $\eta = 0.2 \Delta_0$
$\bigl($$\Delta_0$ is defined such that
the pair potential is
$\Delta({\bf k})=\Delta_0 (\sin k_x \pm i \sin k_y)$
at $T=0$$\bigr)$.
}
\label{fig:2}
\end{figure}
%%%%%%%%%%%%%%%%%%%%%%%%%%%%%%%%%%%

   We show our result for $T_1^{-1}(T)$ in Fig.\ \ref{fig:2}.
   We find that,
even in this case of the anisotropic gap and the anisotropic Ferm surface,
the difference in $T_1^{-1}(T)$ between
the $\sin k_x \pm i \sin k_y$ states
is noticeable
and $T_1^{-1}(T)$ in the $\sin k_x - i \sin k_y$ state
is suppressed in wide $T$ region
in comparison with
that in the $\sin k_x + i \sin k_y$ state.
   This result is
in contrast to a corresponding theoretical result for $T_1^{-1}(T)$
by Takigawa {\it et al.}~\cite{Takigawa02-2}
   In their result,
there is not
such suppression of $T_1^{-1}(T)$ as seen in our Fig.\ \ref{fig:2}.
   While the same anisotropy is taken into account in both calculations,
their result is different from ours.
   A reasonable origin of this difference may be as follows.
   The calculation of $T_1^{-1}(T)$
by Takigawa {\it et al.}~\cite{Takigawa02-2}
is in the quantum limit
($k_{\rm F}\xi \sim 1$)
where the energy spectrum
inside the vortex core is
quantized and
it dominantly determines $T_1^{-1}(T)$
as pointed out by them,~\cite{Takigawa02-2}
while we base our calculation on
the quasiclassical theory relevant in
the opposite limit $k_{\rm F} \xi \gg 1$
where the vortex core spectrum is continuous.
   Now, in the case of the material
Sr$_2$RuO$_4$,~\cite{Sigrist99,Maeno}
the coherence length is not so small
($\xi \sim 660$ \AA~\cite{Akima}),
namely $k_{\rm F} \xi \gg 1$,
and therefore our result based on the quasiclassical theory
is relevant to this material.

% **********************************
\section{SUMMARY}
   Within the framework of the quasiclassical theory of superconductivity,
we numerically calculated
the site-selective
nuclear spin-lattice relaxation rate $T_1^{-1}$
at the vortex center
in the chiral $p$-wave superconductors.
   The case of the tilted magnetic field
and
the case of the anisotropic gap and Fermi surface
were investigated.
   Our result (Fig.\ \ref{fig:1}) can be experimentally observed as a sign of
the chiral pairing state in Sr$_2$RuO$_4$
by applying the magnetic field tilted from
the $z$ crystallographic axis by $\sim 85^{\circ}$.

%%%%%%%%%%%%%%%%%%%%%%%%%%%%%%%%%%%
\section*{ACKNOWLEDGMENTS}

We thank M.\ Ichioka, M.\ Takigawa, M.\ Matsumoto, K.\ Machida,
N.\ Schopohl, and M.\ Sigrist
for helpful discussions.

%%%%%%%%%%%%%%%%%%%%%%%%%%%%%%%%%%%

%%%%%%%%%%%%%%%%%%%%%%%%%%%%%%%%%%%

\end{document}